Marta Tomczak

Uniwersytet Ekonomiczny we Wrocławiu

Anna Ziółkowska

Uniwersytet Ekonomiczny we Wrocławiu

Martyna Rosik

Uniwersytet Ekonomiczny we Wrocławiu

**Rentowność sprzedaży netto przedsiębiorstwa na przykładzie trzech przedsiębiorstw z branży produkcyjnej metalowych wyrobów gotowych (z wyłączeniem maszyn i urządzeń).**

*Working paper*

Streszczenie:

Poniższy raport ma na celu zmierzenie poziomu wskaźnika rentowności sprzedaży netto w trzech przedsiębiorstwach tej samej branży oraz zarekomendowanie odpowiednich działań w zależności od uzyskanych wyników.

<u>Wprowadzenie:</u>

Rentowność sprzedaży netto jest nazywana również zwrotem ze sprzedaży. Informuje ona o udziale zysku netto w wartości sprzedaży. Im niższy jest wskaźnik rentowności, tym większa wartość sprzedaży musi być zrealizowana dla osiągnięcia określonej kwoty zysku. Większa wartość wskaźnika oznacza zatem korzystniejszą sytuację finansową firmy[1].

Wzór na wskaźnik rentowności netto przedstawia się w następująco:

$$RN = \frac{Zysk\ (strata)\ netto}{PN+PPO+PF+ZN} \quad [2] \qquad (1)$$

gdzie:

RN    – rentowność netto

PN    – przychody netto ze sprzedaży towarów i produktów

PPO – pozostałe przychody operacyjne

PF    – przychody finansowe

ZN    – zyski nadzwyczajne

1. <u>Wyznaczanie wskaźnika rentowności sprzedaży netto dla pierwszego przedsiębiorstwa</u>

Tabela 1. Wskaźnik rentowności sprzedaży netto.

| ROK | PRZYCHODY ZE SPRZEDAŻY[3] (w złotych) | WYNIK NETTO (w złotych) | RENTOWNOŚĆ SPRZEDAŻY NETTO (w %) |
|---|---|---|---|
| 2011 | 132 322 600,00 | - 1 540 100,00 | - 1,16 |
| 2010 | 102 596 400,00 | 3 893 300,00 | 3,79 |

*Źródło: Opracowanie własne na podstawie danych ze sprawozdań finansowych*

---

[1] Sierpińska M., Jachna T., *Ocena przedsiębiorstwa według standardów światowych,* Polskie Wydawnictwo Naukowe 2007, str. 198
[2] Wzór zaczerpnięty ze źródła:
Gołaszewski P., *Analiza sprawozdań finansowych*, Fundacja Rozwoju Rachunkowości w Polsce, Łódź 2001, s. 55
[3] Suma przychodów zawarta w mianowniku wzoru numer 1.

Wykres 1. Kształtowanie się wskaźnika rentowności sprzedaży netto w latach 2010-2011.

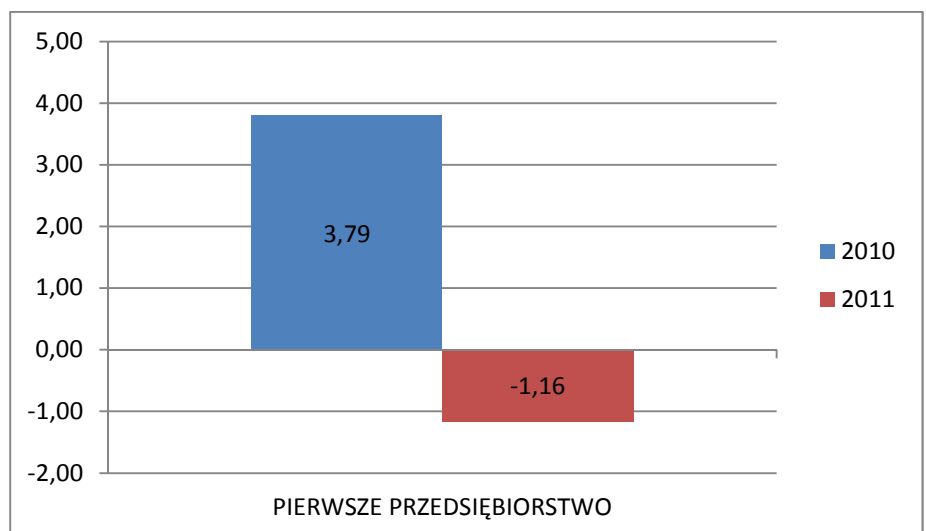

*Źródło: Opracowanie własne na podstawie danych ze sprawozdań finansowych*

Jak wynika z tabeli nr 1 oraz sporządzonego do niej wykresu nr 1, w roku 2011 wskaźnik rentowności sprzedaży pierwszego analizowanego przedsiębiorstwa zmalał o prawie 5 punktów procentowych w stosunku do roku 2010. Spadek ten wynika z tego, iż w 2011 roku wystąpiła strata, natomiast przychód ze sprzedaży przyjął wartość większą niż w roku 2010. Wskaźnik rentowności sprzedaży netto pokazuje, że kolejno w latach 2010 i 2011, wartość zysku/straty netto, przypadających na każdą złotówkę sprzedanych produktów, towarów i materiałów oraz pozostałych składników działalności firmy jest równa odpowiednio 3,79 oraz -1,16. Wiadome jest, że im wyższa jest wartość wskaźnika, tym wyższa jest efektywność dochodów i w 2010 roku dochodowość dla tej firmy była najwyższa spośród badanych lat.

2. <u>Wyznaczanie wskaźnika rentowności sprzedaży netto dla drugiego przedsiębiorstwa</u>

Tabela 2. Wskaźnik rentowności sprzedaży netto.

| ROK | PRZYCHODY ZE SPRZEDAŻY[3] (w złotych) | WYNIK NETTO (w złotych) | RENTOWNOŚĆ SPRZEDAŻY NETTO (w %) |
|---|---|---|---|
| 2011 | 39 201 349,37 | - 2 570 292,48 | - 6,56 |
| 2010 | 35 792 321,65 | 276 819,03 | 0,77 |

*Źródło: Opracowanie własne na podstawie danych ze sprawozdań finansowych*

Wykres 2. Kształtowanie się wskaźnika rentowności sprzedaży netto w latach 2010-2011.

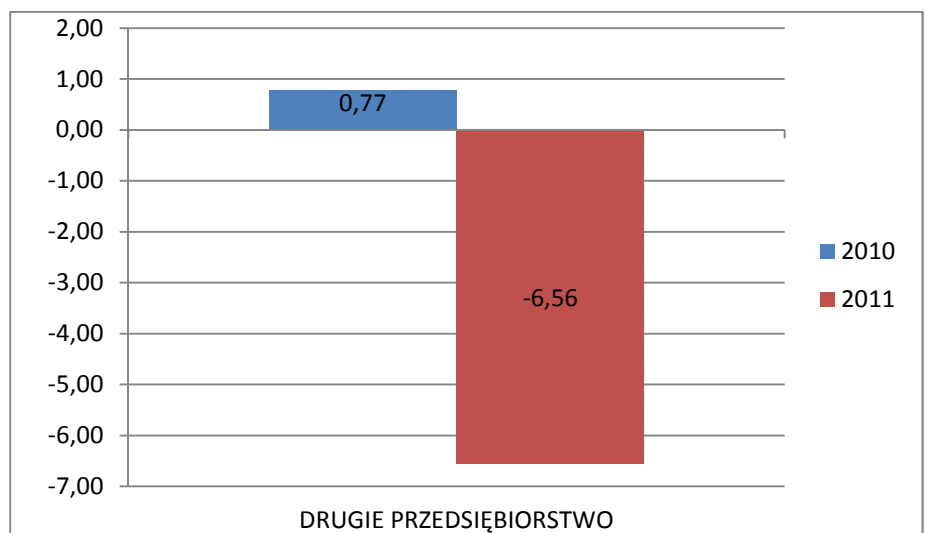

*Źródło: Opracowanie własne na podstawie danych ze sprawozdań finansowych*

Jak wynika z tabeli nr 2 oraz sporządzonego do niej wykresu nr 2, w roku 2011 wskaźnik rentowności sprzedaży drugiego analizowanego przedsiębiorstwa zmalał o ponad 7,3 punktu procentowego w stosunku do roku 2010. Spadek ten wynika z tego, iż w 2011 roku wystąpiła znaczna strata, natomiast przychód ze sprzedaży przyjął wartość większą niż w roku 2010. Wskaźnik rentowności sprzedaży netto pokazuje, że kolejno w latach 2010 i 2011, wartość zysku/straty netto, przypadających na każdą złotówkę sprzedanych produktów, towarów i materiałów oraz pozostałych składników działalności firmy jest równa odpowiednio 0,77 oraz -6,56. Wiadome jest, że im wyższa jest wartość wskaźnika, tym wyższa jest efektywność dochodów i w 2010 roku dochodowość dla tej firmy była najwyższa spośród badanych lat.

3. <u>Wyznaczanie wskaźnika rentowności sprzedaży netto dla trzeciego przedsiębiorstwa</u>

Tabela 3. Wskaźnik rentowności sprzedaży netto.

| ROK | PRZYCHODY ZE SPRZEDAŻY[3] (w złotych) | WYNIK NETTO (w złotych) | RENTOWNOŚĆ SPRZEDAŻY NETTO (w %) |
|---|---|---|---|
| 2011 | 73 305 872,01 | 5 387 044,88 | 7,35 |
| 2010 | 71 424 836,80 | 4 727 609,30 | 6,62 |

*Źródło: Opracowanie własne na podstawie danych ze sprawozdań finansowych*

Wykres 3. Kształtowanie się wskaźnika rentowności sprzedaży netto w latach 2010-2011.

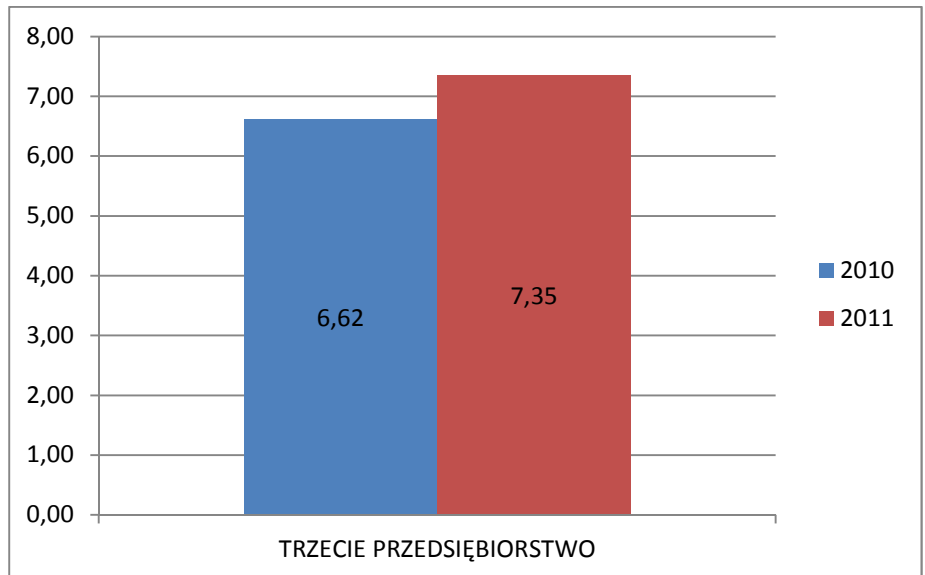

*Źródło: Opracowanie własne na podstawie danych ze sprawozdań finansowych*

Jak wynika z tabeli nr 3 oraz sporządzonego do niej wykresu nr 3, w roku 2011 wskaźnik rentowności sprzedaży trzeciego analizowanego przedsiębiorstwa wzrósł prawie o 0,75 punktów procentowych w stosunku do roku 2010. Wzrost ten wynika z tego, iż w 2011 roku wystąpił wzrost zysku przy jedoczesnym przyjęciu większej niż w 2010 roku wartości przez przychód ze sprzedaży. Wskaźnik rentowności sprzedaży netto pokazuje, że kolejno w latach 2010 i 2011, wartość zysku netto, przypadającego na każdą złotówkę sprzedanych produktów, towarów i materiałów oraz pozostałych składników działalności firmy jest równa odpowiednio 6,62 oraz 7,35. Wiadome jest, że im wyższa jest wartość wskaźnika, tym wyższa jest efektywność dochodów i w 2011 roku dochodowość dla tej firmy była najwyższa spośród badanych lat. Wzrost wskaźnika w tym roku oznaczało, że badana jednostka musiała dla osiągnięcia danej kwoty zysku zrealizować niższą sprzedaż niż wtedy, gdy rentowność sprzedaży była niższa.

Podsumowanie:

Wysoki poziom marży świadczy o tym, że firma potrafi skutecznie pozycjonować swoje strategie cenowe. Dodatkowo utrzymuje koszty swojej działalności na uwięzi[4]. Jednak najlepsza wartość tego wskaźnika uzależniona jest od rodzaju działalności firmy. Przedsiębiorstwa, które mają krótki cykl produkcyjny i możliwość szybkiej sprzedaży produktów, rentowność jest zazwyczaj niższa, co wynika z sytuacji, w której krótki cykl wiąże się z mniejszymi kosztami zamrożenia środków. Ogólnie przyjmuje się, że w dłuższym okresie marża zysku netto nie powinna przynajmniej spadać. Każdy inwestor, stosujący analizę fundamentalną powinien uważnie obserwować poziomy marży w kolejnych okresach, ponieważ może to pomóc w przewidzeniu potencjalnych kłopotów branży.

Zarówno pierwszemu jak również drugiemu analizowanemu przedsiębiorstwu zaleca się wzmożenie czujności i zwiększenie kontroli nad ponoszonymi przez nie kosztami oraz podjęcie prób ich ewentualnego zmniejszenia. Natomiast trzeciemu analizowanemu przedsiębiorstwu zaleca się kontynuację działalności w normalnym trybie jej funkcjonowania.

---

[4] Biedrzycki P., *Marża zysku netto (wskaźnik rentowności sprzedaży)*, 2008, dostępny na: http://sindicator.net/baza_wiedzy/wskazniki_rentownosci_i_oceny_perspektyw_rozwojowych/marza_zysku_n etto_wskaznik_rentowno

Literatura:

1. Michalski Grzegorz, Ocena finansowa kontrahenta na podstawie sprawozdań finansowych (Financial Analysis in the Firm. A Value-Based Liquidity Framework), oddk, Gdańsk 2008.

2. Grzegorz Michalski (2012), FINANCIAL LIQUIDITY MANAGEMENT IN RELATION TO RISK SENSITIVITY: POLISH FIRMS CASE, Proceedings of the International Conference Quantitative Methods in Economics (Multiple Criteria Decision Making Xvi), 141-160

3. Michalski Grzegorz, Strategie finansowe przedsiębiorstw (Entrepreneurial financial strattegies), oddk, Gdańsk 2009.

4. Michalski, Grzegorz Marek, Wprowadzenie do zarządzania finansami przedsiębiorstw, (Introduction to Entrepreneurial Financial Management), Available at SSRN: http://ssrn.com/abstract=1934041 or http://dx.doi.org/10.2139/ssrn.1934041

5. Michalski Grzegorz (2007), Portfolio Management Approach in Trade Credit Decision Making, Romanian Journal of Economic Forecasting, Vol. 3, pp. 42-53, 2007. Available at SSRN: http://ssrn.com/abstract=1081269

6. Michalski Grzegorz (2008), Operational risk in current assets investment decisions: Portfolio management approach in accounts receivable, Agricultural Economics–Zemedelska Ekonomika, 54, 1, 12–19

7. Michalski Grzegorz (2008), Corporate inventory management with value maximization in view, Agricultural Economics-Zemedelska Ekonomika, 54, 5, 187-192.

8. Michalski Grzegorz (2009), Inventory management optimization as part of operational risk management, Economic Computation and Economic Cybernetics Studies and Research, 43, 4, 213-222.

9. Michalski Grzegorz (2011), Financial Analysis in the Enterprise: A Value-Based Liquidity Framework. Available at SSRN: http://ssrn.com/abstract=1839367, 177-262.

10. Michalski Grzegorz (2007), Portfolio management approach in trade credit decision making, Romanian Journal of Economic Forecasting, 8, 3, 42-53.

11. Michalski Grzegorz (2008), Value-based inventory management, Romanian Journal of Economic Forecasting, 9, 1, 82-90.

12. Michalski Grzegorz (2012), Financial liquidity management in relation to risk sensitivity: Polish enterprises case, Quantitative Methods in Economics, Vydavatelstvo EKONOM, Bratislava, 141-160.

13. Michalski Grzegorz (2008), Decreasing operating risk in accounts receivable mangement: influence of the factoring on the Enterprise value, [in] Culik, M., Managing and Modelling of Financial Risk, 130-137.

14. Michalski Grzegorz (2010), Planning optimal from the Enterprise value creation perspective. Levels of operating cash investment, Romanian Journal of Economic Forecasting, vol: 13 iss: 1 pp.198-214.
15. Polak Petr, Robertson, D. C. and Lind, M. (2011), The New Role of the Corporate Treasurer: Emerging Trends in Response to the Financial Crisis (December 12, 2011). International Research Journal of Finance and Economics, No. 78, Available at SSRN: http://ssrn.com/abstract=1971158
16. Soltes Vincent (2012), Paradigms of Changes in the 21th Century - Quest for Configurations in Mosaic, Ekonomicky Casopis, v.60 is.4 pp. 428-429.
17. Soltes Vincent (2011), The Application of the Long and Short Combo Option Strategies in the Building of Structured Products, 10th International Conference of Liberec Economic Forum, Liberec.
18. Zmeskal Zdenek, Dluhosova Dana (2009), Company Financial Performance Prediction on Economic Value Added Measure by Simulation Methodology, 27th International Conference on Mathematical Methods in Economics, Mathematical Methods in Economics, 352-358.
19. Polak Petr, Sirpal R., Hamdan M. (2012), Post-Crisis Emerging Role of the Treasurer, European Journal of Scientific Research, 86, 3, 319-339
20. Kresta A.; Tichy Tomas (2012), International Equity Portfolio Risk Modeling: The Case of the NIG Model and Ordinary Copula Functions, FINANCE A UVER-CZECH JOURNAL OF ECONOMICS AND FINANCE 62, 2, 141-161.
21. Kopa Milos, D'Ecclesia RL, Tichy Tomas (2012), Financial Modeling, FINANCE A UVER-CZECH JOURNAL OF ECONOMICS AND FINANCE, 62, 2, 104-105.
22. Michalski, Grzegorz Marek, Value-Based Inventory Management, Value-Based Inventory Management, Journal of Economic Forecasting, 9/1, 82-90, 2008. Available at SSRN: http://ssrn.com/abstract=1081276 or http://dx.doi.org/10.2139/ssrn.1081276
23. Dluhosova Dana, et. al., 2006, Finanční řizeni a rozhodovani podniku: analyza, investovani, oceňovani, riziko, flexibilita, Ekopress, Prague.
24. Soltes Vincent, 2004, Duration of coupon bonds as a criterion of the price sensibility of bonds with regards to the change of interest rates (Duracia kuponovej obligacie ako kriterium cenovej citlivosti obligacie vzhľadom na zmenu urokovych sadzieb in Slovak), EKONOMICKY CASOPIS, 52/2004(1), pp. 108-114.
25. Michalski, Grzegorz Marek, Factoring and the Firm Value (May 17, 2008). FACTA UNIVERSITATIS Series: Economics and Organization, Vol. 5, No. 1, pp. 31-38, 2008. Available at SSRN: http://ssrn.com/abstract=1844306
26. Michalski, Grzegorz Marek, Crisis Caused Changes in Intrinsic Liquidity Value in Non-Profit Institutions (December 14, 2012). Equilibrium. Quarterly Journal of Economics and Economic

Policy, 2012, Volume 7, Issue 2. Available at SSRN: http://ssrn.com/abstract=2189488 or http://dx.doi.org/10.2139/ssrn.2189488

27. Michalski, Grzegorz Marek, Płynność finansowa w małych i średnich przedsiębiorstwach (Financial Liquidity Management in Small and Medium Enterprises) (2013). Płynność Finansowa w Malych i Srednich Przedsiebiorstwach, PWN, 2013. Available at SSRN: http://ssrn.com/abstract=2214715

28. Michalski, Grzegorz Marek, Accounts Receivable Management in Nonprofit Organizations (Zarządzanie należnościami w organizacjach nonprofit), 2012, Zeszyty Teoretyczne Rachunkowości 2012(68(124)):83-96. ICID: 1031935, Available at SSRN: http://ssrn.com/abstract=2193352 or http://dx.doi.org/10.2139/ssrn.2193352

29. Tichy, T. (2011), Levy Processes in Finance: Selected applications with theoretical background. SAEI, vol. 9. Ostrava: VŠB-TU Ostrava, ISBN 978-80-248-2536-6.

30. Tichy, T. (2008), Lattice models – Pricing and Hedging at (In)complete Markets. VŠB-TU Ostrava, ISBN 978-80-248-1703-3

31. M. Kopa & T. Tichy (2012), Concordance measures and second order stochastic dominance – portfolio efficiency analysis, E & M Economics and Management 4, 110-120

32. Franków N., Czajkowska A., Zarządzanie wartością przedsiębiorstwa na przykładzie przedsiębiorstw z branży produkcyjnej maszyn i samochodów. (March 22, 2013). Available at SSRN: http://ssrn.com/abstract=2238066 or http://dx.doi.org/10.2139/ssrn.2238066

33. Jaskulska K., *Ocena kondycji finansowej przedsiębiorstwa „Lentex" w latach 2008-2010,* Społeczna Wyższa Szkoła Przedsiębiorczości i Zarządzania w Łodzi, Łódź 2011

34. Juja T., *Analiza finansowa i opodatkowanie przedsiębiorstw. Studium przypadków,* Wydawnictwo Forum Naukowe, Poznań 2007, 123-130

35. Nowak E., *Analiza sprawozdań finansowych,* Polskie Wydawnictwo Ekonomiczne, Warszawa 2008, 175-198

36. Gołaszewski P., Urbanek P. Walińska E., *Analiza sprawozdań finansowych*, Fundacja Rozwoju Rachunkowości w Polsce, Łódź 2001, 46-49

37. Wędzki D, *Analiza wskaźnikowa sprawozdania finansowego,* Wolters Kluwer, Kraków 2006, 492-531

38. Kowalak R., *Ocena kondycji finansowej przedsiębiorstwa w badaniu zagrożenia upadłości,* ODDK, Gdańsk 2008, 127-136

39. Ćwiąkała-Matys A., Nowak W., *Zarys metodologiczny analizy finansowej,* Wydawnictwo Uniwersytety Wrocławskiego, Wrocław 2005, 111-113

40. Rutkowski A., *Zarządzanie finansami,* Polskie wydawnictwo Ekonomiczne, Warszawa 2007, 94-95

41. Dębski W., *Teoretyczne i praktyczne aspekty zarządzania finansami przedsiębiorstwa,* Wydawnictwo Naukowe PWN, Warszawa 2005, 96-98
42. Mohr A., *Zarządzanie finansami. Co mówią liczby,* Wydawnictwo Helion, 2006, 52-53
43. Sierpińska M., Jachna T., *Ocena przedsiębiorstwa według standardów światowych,* Polskie Wydawnictwo Naukowe 2007, str. 198
44. Biedrzycki P., *Marża zysku netto (wskaźnik rentowności sprzedaży),* 2008, dostępny na: http://sindicator.net/baza_wiedzy/wskazniki_rentownosci_i_oceny_perspektyw_rozwojowych/marza_zysku_netto_wskaznik_rentowno